# Manually Acquiring Targets from Multiple Viewpoints Using Video Feedback

Bailey Ramesh, Anna Konstant, Pragathi Praveena, Emmanuel Senft, Michael

Gleicher, Bilge Mutlu,

Michael Zinn, and Robert G. Radwin*

University of Wisconsin-Madison, Madison, WI USA

*Corresponding Author

Address correspondence to:
Robert G. Radwin, PhD
Department of Industrial and Systems Engineering
1550 Engineering Drive
Madison, WI 53706
e-mail: rradwin@wisc.edu

Running Head:  Manually Acquiring Targets Using Video

# Abstract

**Objective:** The effect of camera viewpoint was studied when performing visually obstructed psychomotor targeting tasks.

**Background:** Previous research in laparoscopy and robotic teleoperation found that complex perceptual-motor adaptations associated with misaligned viewpoints corresponded to degraded performance in manipulation. Because optimal camera positioning is often unavailable in restricted environments, alternative viewpoints that might mitigate performance effects are not obvious.

**Methods:** A virtual keyboard-controlled targeting task was remotely distributed to workers of Amazon Mechanical Turk. The experiment was performed by 192 subjects for a static viewpoint with independent parameters of target direction, Fitts' law index of difficulty, viewpoint azimuthal angle (AA), and viewpoint polar angle (PA). A dynamic viewpoint experiment was also performed by 112 subjects in which the viewpoint AA changed after every trial.

**Results:** AA and target direction had significant effects on performance for the static viewpoint experiment. Movement time and travel distance increased while AA increased until there was a discrete improvement in performance for 180º. Increasing AA from 225º to 315º linearly decreased movement time and distance. There were significant main effects of current AA and magnitude of transition for the dynamic viewpoint experiment. Orthogonal direction and no-change viewpoint transitions least affected performance.

**Conclusions:** Viewpoint selection should aim to minimize associated rotations within the manipulation plane when performing targeting tasks whether implementing a static or dynamic viewing solution. Because PA rotations had negligible performance effects, PA adjustments may extend the space of viable viewpoints.

**Applications:** These results can inform viewpoint-selection for visual feedback during psychomotor tasks.

## Précis

Subjects performed a planar targeting task in a virtual environment for a fixed control frame but for varying viewpoints. Viewpoint was an important factor in targeting performance for both static and dynamic views. These results inform viewpoint selection for visual feedback in real-world manual tasks.

# Introduction

*Objective*

Obstructed visual access in aviation manufacturing presents numerous ergonomic challenges while performing assembly operations (Menegon & Fischer, 2012; Mueller et al., 2017; Mueller et al., 2019). In many cases, workers need to peer into confined spaces to reach inside and access work pieces, perform tasks solely by feel, or use a mirror to see around obstacles. Oftentimes, this necessitates awkward neck and shoulder postures, resulting in adverse effects such as discomfort, muscle fatigue, and muscle strain (Beuß et al., 2019). The obstacles associated with limited visual access often result in impaired task performance and completion time, which exacerbates these negative ergonomic outcomes.

A potential solution to this problem is to provide visual feedback via a mobile camera that can be placed inside the workspace, relaying the visual field of the task back to the worker. Although this may seem like a simple solution, camera placement is not always intuitive. Aviation manufacturing workers commonly have to perform tasks in confined spaces, and because of this, the space for available camera viewpoints is often limited by geometric constraints of the aircraft structure. In addition, certain camera viewpoints may hinder task performance due to perceptual-motor mismatch resulting from an inconsistency between the performed hand or tool movement and the visual feedback from the camera. While extensive work has gone into minimizing such perceptual-motor misalignments in teleoperation and robotic surgery, similar questions remain regarding camera pose selection when performing manual tasks with visual assistance in other settings. Teleoperation affords the ability to realign control frame of reference with the visual display, but it is not possible to computationally realign hand and arm movements with visual viewpoints because the monitor and the controller are decoupled (Hiatt & Simmons, 2006). Thus, the operator must perform perceptual-motor adaptations to reconcile movements displayed in conjunction with their body orientation (Cunningham, 1989). This study aims to build on prior work in viewpoint location in an effort to better inform viewpoint selection given a limited space of candidate viewpoints and varying worker-work piece relationships.

*Background*

In certain applications, video is used to provide visual feedback when the line of sight is obstructed. An example is the use of video in medical laparoscopic procedures (Gerges et al., 2006). In these operations, the laparoscope is used to relay a visual field of the abdomen back to the surgeon to inform their operation or diagnosis. Prior work on camera positioning in this field has focused on the effects of different visual feedback presentation variables in an effort to maximize surgical performance. Wentink et al. (2000) determined that the orientation of the perceived endoscopic tool had a direct effect on performance. They found that by eliminating misorientations between the movements of the displayed image and that of the actual surgical tool, both execution time and mental effort were significantly improved.

Wentink et al. (2002) investigated the importance of kinematic effects in endoscopic surgery. These effects arise from the pivoting motion that results from operating through an incision, whereby the movements of the instrument tip appear mirrored and scaled relative to the movements of the external tool handle. In that experiment, normal endoscopic manipulation was compared against a condition in which the scaling and mirror effects were controlled by transforming the displayed visual field. Upon controlling for these effects, they found that the efficiency of endoscopic manipulation could be significantly improved. This was indicated by an improvement in mean task completion time which demonstrates that these negative kinematic effects degrade eye-hand coordination. Task completion time was also improved in Fernandez and Bootsma (2004). Linear and logistic mapping was used in a target aiming task with a pointer and stylus. The results showed that the orientation had an effect on the movement time, but not on the kinematic pattern.

In addition to understanding the importance of orientation and kinematic effects, Hanna et al. (1998) provided evidence of the importance of display location in endoscopic surgery. In that study, subjects were asked to perform an endoscopic knotting task with varying monitor locations. The horizontal monitor position was either in front, to the left, or to the right of the operators' hands. The vertical location conditions were an eye-level and hand-level monitor location. Ultimately, the subjects were able to perform the knotting task faster and with a higher performance score when the monitor was located in front of the subject and at hand level. Under these conditions, the misalignment between the operator's head stance, workspace, and the monitor were minimized.

The above-mentioned improvements can be greatly attributed to stimulus-response compatibility in which perceived movements are more congruent with expected movements (Cho & Proctor, 2003). Inconsistencies between perceived and expected movements evoke perceptual-motor adaptation to reconcile the mismatch (Redding et al., 2005). During adaptation, it appears that a person gradually updates their internal mapping that transforms visual information into motor commands (Cunningham, 1989). However, these adaptations can degrade task performance due to increased mental workload (Klein et al., 2005). The need to perform these adaptations hinders performance in other modes of manual control as well, such as in teleoperation of master-slave robotic systems. DeJong et al. (2004) showed that the mental workload in master-slave teleoperation can be reduced by eliminating or minimizing control rotations, view rotations, and control translations. Of these, reducing control rotations has the most significant benefit to performance (DeJong et al., 2011). These control rotations can be eliminated by adjusting manipulandum, display, robot, and camera coordinate frames such that the rotation from the camera frame to the robot frame is equal to the rotation from the display frame to the manipulandum frame. By satisfying this equation, a more compatible mapping from the local control environment to the remote environment can be ensured.

In addition to the geometric restrictions imposed in the manufacturing space, certain tasks may also present a dynamic workspace as the task is being completed. As such, the space of available viewpoints may change with time. Therefore, to implement a superior viewing solution of this nature, we first need to understand how the underlying mechanisms of common manual tasks are affected by viewpoint in two regards: from a static camera and from a changing camera angle.

This study aims to address these questions by presenting two experiments designed to learn which viewpoints least hinder task performance due to perceptual-motor distortions between expected and achieved hand or tool movements. In the first, participants were tasked with performing a simple target acquisition task from a stationary viewpoint. In the second experiment, subjects performed the same task but from a changed and discretely cut viewpoint. It was hypothesized that certain viewpoint angles and target directions will impose a significantly greater mental workload during task completion, and this will be reflected in performance of the task, including movement time and movement distance. It is also hypothesized that performance at a particular viewpoint is dependent on recent camera pose azimuthal angle selection when effectuating a dynamic perspective solution. Through testing these propositions, we may yield results that can be used to make recommendations for camera placement given a constrained set of candidate viewpoints in a variety of work environments.

# Methods

*Research Environment*

This research was conducted during the outbreak of the coronavirus disease 2019 pandemic which placed significant limitations on research involving physical interactions with human participants (Servick et al., 2020). Previously, viewpoint manipulation effects would have been evaluated directly via a psychomotor task where parameters could be physically altered, and task performance could be measured in a physical experimental set-up. Due to the strict limitations for preventing virus transmission, the study shifted to a remote virtual setting to eliminate physical contact with subjects. The Unity game engine was used to develop the following virtual experiment which was then uploaded to a server to access participants remotely without the need to enter the laboratory.

*Participants*

Participants were recruited, with informed consent and IRB approval, from the Amazon Mechanical Turk crowdsourcing workforce via the internet. Following successful completion of the experiment, they were compensated for their time through Mechanical Turk. A total of 192 participants (118 male, 74 female) for Experiment 1, and 112 subjects (70 male, 42 female) for Experiment 2 were recruited. Inclusion was limited to English-speaking workers located geographically in the United States. Physical requirements included normal or corrected vision and the absence of any movement impairments in the upper limbs.

Prior to the experiment, subjects completed a pre-screening survey to exclude participants with inadequate computer hardware. Required hardware included a laptop with an attached keyboard and arrow keys arranged in one of three intuitive spatial layouts (Figure 1).

This was done to ensure a similar physical relationship between controls and display across subjects and to avoid the effect of different control mappings for subjects with irregular arrow key arrangements. There was no restriction on laptop size or laptop set up. Subjects could orient and arrange their laptop in a way that was most comfortable for them.

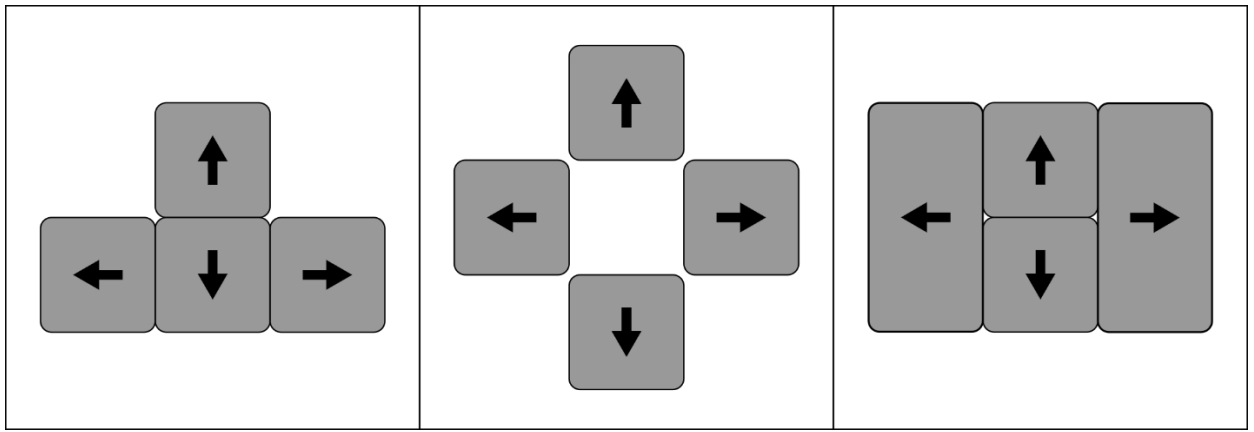

**Figure 1.** Subjects were asked to select from one of the above array key layouts or select "other" if their layout was not shown. Those who selected "other" were excluded from the study.

### Experiment 1: Target Acquisition for a Static Viewpoint

*Experimental Task*

A virtual Fitts' law targeting task was developed in a 3D virtual environment using the Unity game engine to assess the effect of viewpoint on target acquisition. For this task, subjects were asked to direct a red cursor into a green target area as quickly as possible using the computer keyboard arrow keys (Figure 2). The participant viewpoint of the game environment was changed to quantify its effects on the dependent variables: movement time and movement path distance. The viewpoint consisted of a polar viewpoint angle (PA) and azimuthal viewpoint angle (AA). The PA was the height above or below the horizontal plane. The AA was the angle of the rotated camera. Keyboard controls were selected because their physical characteristics were less variable than those of computer mice or track pads.

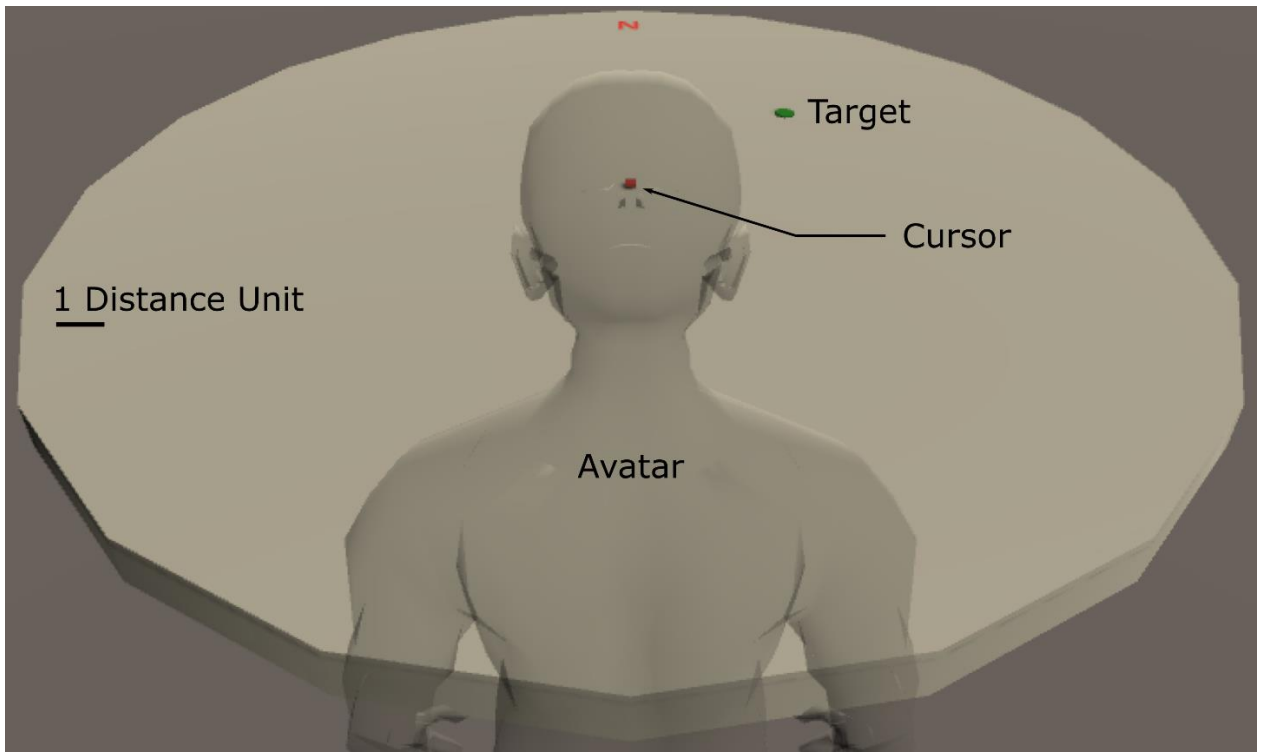

**Figure 2.** The Unity virtual environment. In Experiment 1, the subject was asked to direct the red cursor into the green target area using the computer arrow keys. Cursor movements were made with respect to the avatar shown in the display. In addition to the avatar, a red "N" was displayed to signify "North," or the direction of movement aligned with the forward arrow key. These visual aids served as control references when the perspective was changed.

A trial is defined as each target acquisition task. At the beginning of each trial, the cursor was initialized at the center of the virtual environment. Task timing began after the first movement of the cursor and ended upon entering the target area. Once entered, the target color changed from dark green to light green to inform the subject that the cursor was admitted. The cursor then had to remain inside the target zone for 0.5 s in order to be acquired. If the cursor left the target area before the 0.5 s elapsed, the trial continued until successful target acquisition. Movement path distance was defined as the total path length traversed when acquiring the target.

Movement direction was aligned with the arrow keys with respect to the virtual avatar's position. A red "N" (i.e. North) was displayed to indicate the forward direction, or the direction traveled when the up arrow key was pressed, and inform the subject of the changed perspective in the virtual space. Likewise, the down arrow key moved the cursor towards the avatar, and the left and right arrow keys resulted in the corresponding cursor movement with respect to the perspective of the avatar. Cursor movement was limited to discrete steps in orthogonal directions in alignment with the keyboard arrows. Movement was proportional to the rate of key tapping where the resulting cursor step displacement increased with greater keying rates. Sustained key presses did not result in repeated steps and movements were limited to horizontal and vertical increments. A release between each key press was required in order to make the keystrokes needed for straight and diagonal directions more equivalent. Movement was controlled by the Unity *AddForce* function acting upon the cursor *Rigidbody*. This force was applied in *Velocity Change* mode, the surface friction coefficient was set to zero, and the mass and drag of the *Rigidbody* cursor were set to 0.5 and 15, respectively. The magnitude of each step was inversely related to the rate of key tapping to better allow the subject to control the speed of movement (Figure 3). All cursor values are provided in Unity force units. The parameters of this control scheme were selected based on their ability to allow movement conformant to Fitts' law from a preliminary experiment.

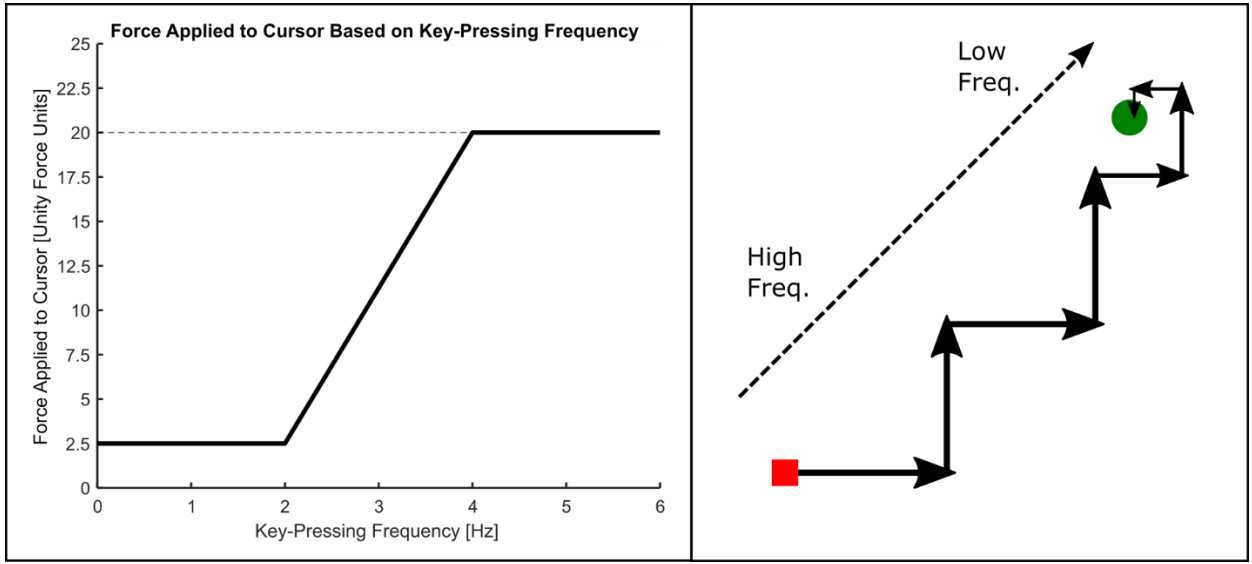

**Figure 3.** The force applied to the cursor in Unity is shown for a range of key-tapping rates and offers linear force modulation between 2 and 4 Hz (left). This allows larger movement amplitudes during the ballistic movement phase and smaller movement amplitudes for fine adjustment into the target (right).

Camera position was defined using a spherical coordinate system (Figure 4a). The polar angle φ was the angle between the Z-axis normal of the work surface and the line segment OC. The azimuthal angle θ was the angle between the positive X-axis and the projection of OC onto the work surface. A negative angle of φ will move the camera to the other side of the z axis, leading to an inverted image as the camera is facing upside down (as shown in Figure 4b). . Because we were primarily interested in the effect of camera rotations in manipulation, the radial distance was held constant at 25 Unity API units throughout the experiment. As such, the camera position was defined solely by the azimuthal and polar angles. Three example camera positions and their respective views are shown in Figure 4b.

Following the Mechanical Turk pre-screening survey, subjects were provided a link to the experiment website. They were then briefed on the experiment through an interactive tutorial prior to data collection. In this tutorial, subjects were instructed to hit the green target as fast as they could while maintaining accuracy. They were informed about the purpose of the avatar and red "N" landmarks in providing cursor movement references. Following this, they were allowed to practice moving the cursor in the absence of a target until they indicated they were ready to proceed. This was done to help subjects become familiar with the control scheme for cursor movement. Subjects were shown that the timer would begin upon first movement, so that they would not be rushed to begin a trial immediately after loading. Before starting the experiment, they practiced the task by performing three trials at the view aligned with the avatar's eyes (0°, 60°) and three trials at an alternative view (270°, 60°).

*Experimental Design*

A full factorial experimental design was utilized with four independent variables: eight AAs, three PAs, eight target directions (TD), and three indices of difficulty (ID). The levels of each of these variables are shown below in Table 1. The three IDs of 2, 3.5, and 5 were assigned target amplitudes of 3, 6, and 9 Unity API units and target diameters of 1.5, 1.06, and 0.5625 Unity API units, respectively. Because cursor movement was limited to orthogonal steps, target amplitudes were defined using the Manhattan distance to ensure diagonal targets could be captured in the same number of keystrokes as horizontal and vertical targets. Each subject was presented all 24 TD x ID combinations for one of the 24 PA x AA viewpoint combinations in a random order. The PA x AA angles were counterbalanced evenly to provide eight subject replications for each viewpoint. In this way, a between-subjects design was used that confounded viewpoint effects among subjects.

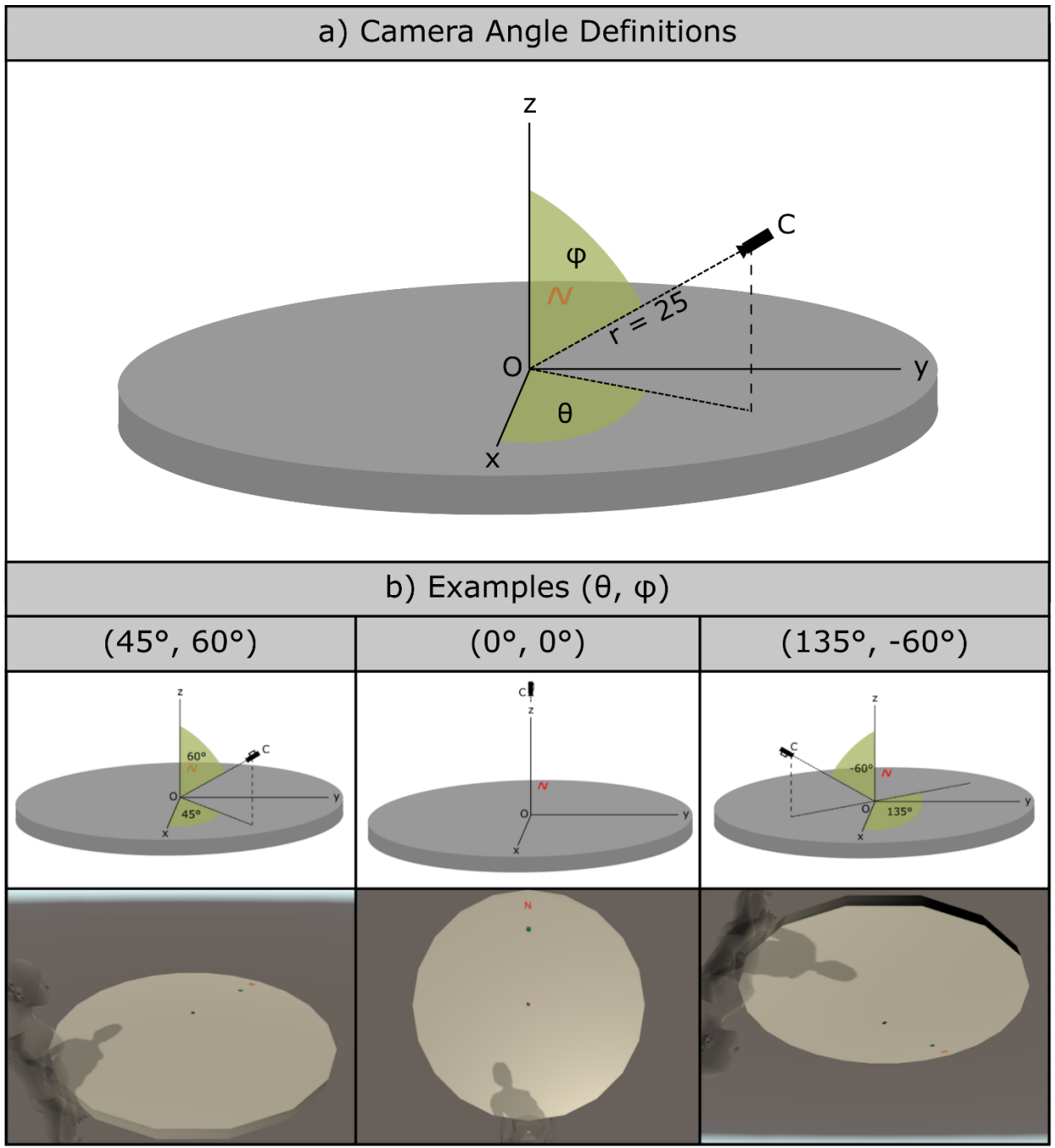

**Figure 4.** a) The camera position is defined using the spherical coordinates (r, θ, φ) where r is the radial distance to the center of the cylindrical surface, θ is the azimuthal angle, and φ is the polar angle. The avatar is aligned with the positive x-axis. b) Three example viewpoints and their corresponding azimuthal and polar angle definitions. The upper panels illustrate the camera positions, and the lower panels depict the corresponding displays. The main part of the avatar visible is the shadow on the table.

Table 1. Independent Variables and Levels

| Variable | Levels |
| --- | --- |
| Viewpoint Azimuthal Angle (AA) | 0°, 45°, 90°, 135°, 180°, 225°, 270°, 315° |
| Viewpoint Polar Angle (PA) | -60°, 0°, 60° |
| Target Direction (TD) | 0°, 45°, 90°, 135°, 180°, 225°, 270°, 315° |
| Index of Difficulty (ID) | 2, 3.5, 5 |

      For the purposes of this experiment, a set included 24 target acquisitions: one acquisition for each of the three IDs in each of the eight target directions (Figure 5). Target direction was defined such that the 0° direction pointed towards the avatar, the 180° direction pointed towards the red "N", and the angle increased in the counterclockwise direction when viewed from above. Subjects first performed seven randomized sets of target acquisitions in total: two at the eye view perspective and then five at their assigned alternative viewpoint. Between sets, subjects were presented with an intermission screen to allow a break if desired. The first two sets from the eye view perspective, during the experiment, served as a training period prior to presenting the alternative viewpoint to further allow subjects practice with the movement controls. From a preliminary study, target acquisition time was observed to level off after approximately twenty acquisitions. In an effort to omit learning variations among subjects, the first of the five sets of data at the alternative viewpoint was withdrawn as practice. The subsequent analysis was performed on the remaining four sets of data for each subject in which performance was more stabilized.

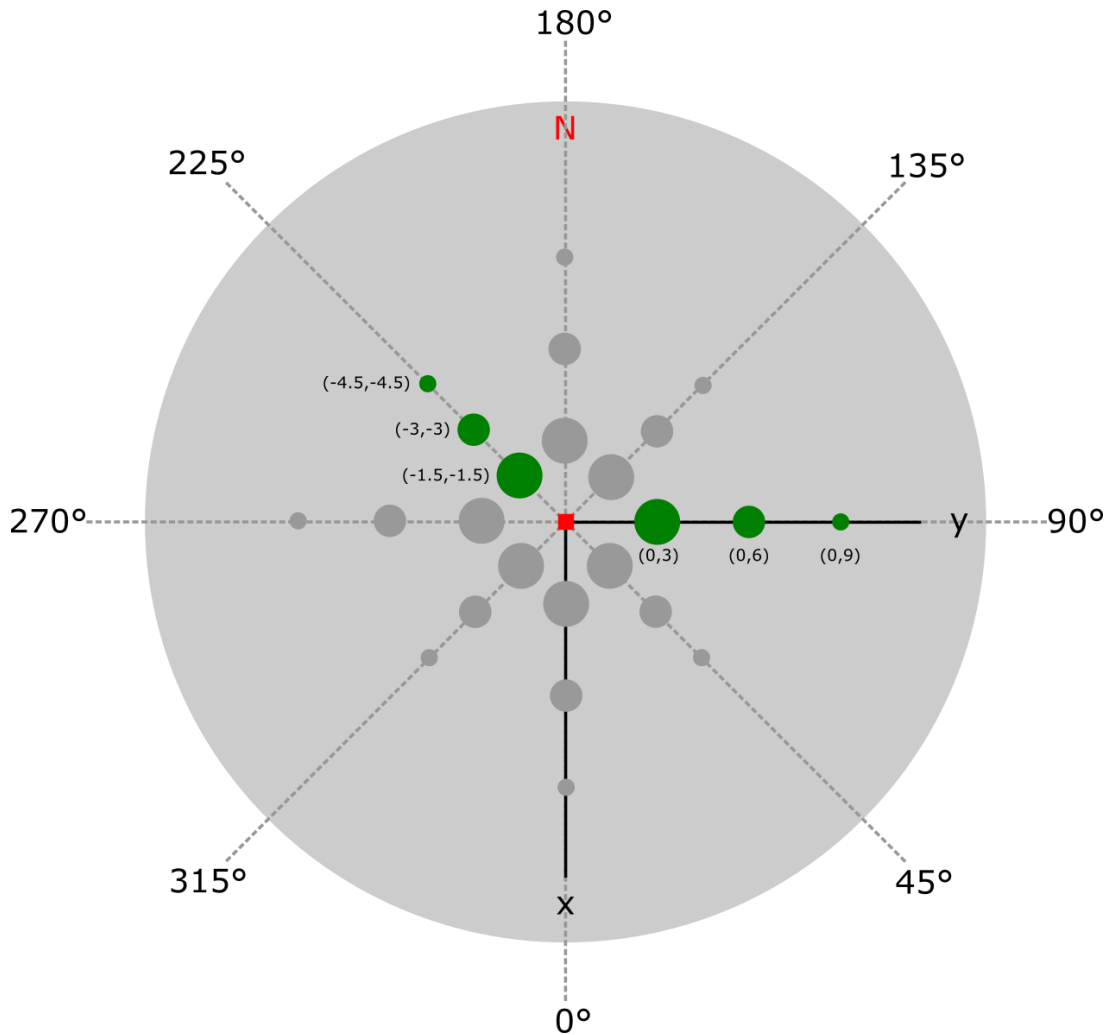

**Figure 5.** Target directions, distances, and diameters. The target direction angle is always given with respect to the position of the avatar, starting at 0° and increasing in the counterclockwise direction. Six example target configurations are shown in green: three in the diagonal direction 225° and three in the horizontal direction 90°. The corresponding coordinates are given in Unity distance units using the previously defined coordinate system. The remaining 18 target configurations are shown in grey with coordinates omitted for clarity. All 24 targets illustrate the 24 trials that comprise a set.

*Data Analysis*

Potential outliers in the data were identified by task times and intervals of time between cursor movements greater than five standard deviations. Within these outliers, the keying patterns and intervals of time between cursor movements were checked to ensure subjects were following the instructions. Three trials were removed (.001%) from the data set because the trial contained an interval greater than or equal to 15 seconds. A linear mixed model ANOVA was used to test the learning effects on the dependent variables of movement time and movement distance. The independent variables used in the model were subjects and set number. Subjects are included as a random intercept in the model to account for the variance caused by different subjects in the experiment.

The R Project for Statistical Computing was used for data analysis. A linear mixed model tested the effects of the between-subject factors (AA and PA) and the within-subject factor (target direction) on movement time and distance. The independent variables in the model were polar angle, azimuthal angle, target direction, and subjects. Subjects were treated as a random intercept in the model, to account for subject variance. Transformations were applied to the dependent variables for stabilizing the variance when necessary. The residual plots were examined for normality and the best transformation was chosen (Montgomery, 2012). A reciprocal transformation was used for movement time and a log transformation was used for distance. All post hoc analyses were performed using Tukey pairwise comparisons ($α = .05$).

## Experiment 2: Target Acquisition for a Changing Viewpoint

*Experimental Task*

In the second experiment, subjects were presented the same task of directing the cursor into the green target area. Instead of acquiring targets from a static viewpoint, however, the viewpoint was instantly changed after each trial by making a discrete cut from one viewpoint to the next. A discrete cut was chosen based on the assumption two cameras would be used to complete the task. Cursor movement control settings remained the same as set in Experiment 1. The same dependent variables were used in Experiment 2, in addition to reaction time. Reaction time was defined as the time between the onset of the target and the onset of the first movement.

*Experimental Design*

To minimize the effect of target configuration in this experiment, only diagonal target directions were used. The target direction was selected randomly for each trial from the four possible directions. The target ID was set to 3.5 throughout the experiment with the same parameters as in Experiment 1. Subjects were provided with the same pre-experiment tutorial as given in Experiment 1.

The polar camera angle φ was held constant at 60° throughout Experiment 2. The viewpoint subset of interest was composed of the eight camera locations defined by the eight azimuthal angles 0°, 45°, 90°, 135°, 180°, 225°, 270°, and 315°. For the initial trial of a set, the viewpoint was set randomly to one of these eight perspectives. During the following 64 trials, the viewpoint shuffled randomly through all possible transitions among the eight angles (Figure 6). This included the eight null transitions, or transitions from one viewpoint to itself which act as a control transition for each viewpoint.

Each transition was defined using the two independent parameters, the previous AA and current AA. A within-subjects full factorial design approach was used to analyze the effects of each of these variables on the dependent performance parameters, movement time and movement distance. The 64 trials comprising a set were administered three times, again with intermissions provided in between. The first set was withdrawn as a practice set, and the following data analysis was performed on the remaining two sets of trials for each subject.

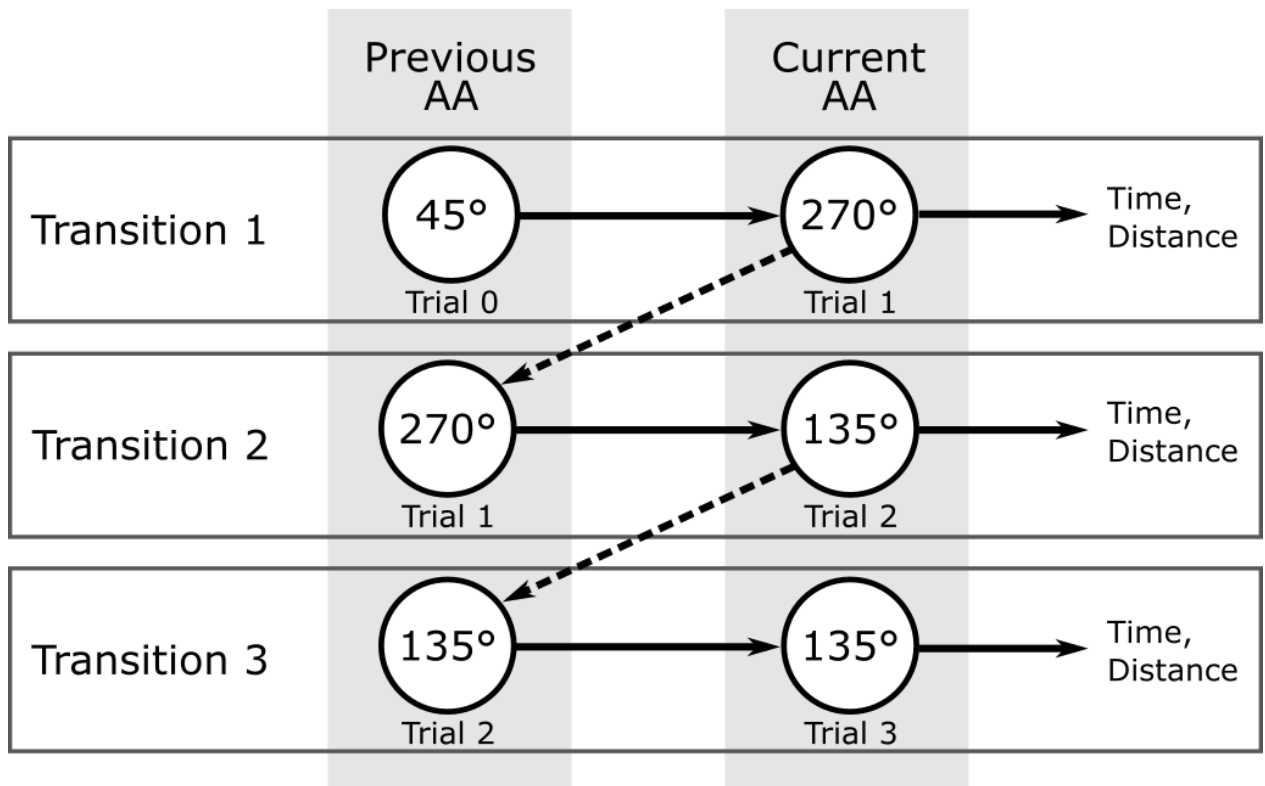

**Figure 6.** Shown above are the first three transitions for an example set of trials. In Trial 0, the viewpoint is randomly chosen as 45°. Following completion of this trial, the scene reloads with a random AA of 270° for Trial 1. The time and distance data for this trial is then representative of the 45° to 270° transition. After Trial 1, the scene reloads with another random AA of 135°, and the performance data from Trial 2 is classified as the 270° to 135° transition. The last transition, Transition 3, is an example of a no-change transition, or successive trial at the same AA.

*Data Analysis*

Outliers in time between the initial movements and subsequent movements were identified, using the same criteria of 15 seconds. A total of 30 trials were removed from the analysis (.138%). In addition, the learning effects were studied between the sets, using a linear mixed effects model with set number as a fixed effect and subjects as a random intercept for subject variance. Transformations were applied to the dependent variables for stabilizing the variance when necessary. The residual plots were examined for normality and the best transformation was chosen (Montgomery, 2012). The movement time data was stabilized with a reciprocal square root transformation. The reaction time was stabilized by adding a constant of 1 to the dependent variable and applying a log transformation. Further, the distance data was stabilized using a log transformation. A linear mixed model ANOVA was conducted to identify the effect of previous AA and current AA on performance. The dependent variable time passed the Bartlett Test for Homogeneity (p = .273) but did not pass for movement distance (p < .0001). An additional linear mixed effect model was conducted to study the magnitude of transition and current AA on the dependent variables. Subjects were considered a random intercept in the model to account for the differences between subjects. The same approach was used to examine the residual plots in Experiment 2 and the best transformations were chosen. The residual plots of the transformed dependent variables were all analyzed, and no trends were observed. All post hoc analyses were performed using a Tukey pairwise comparison (α = .05) to determine the differences in independent variables.

# Results

## Experiment 1: Target Acquisition for a Static Viewpoint

*Movement Time*

There was a significant effect of learning among the five sets (F(4,22844) = 81.43, p < .0001) where the movement time decreased as more sets were completed. The greatest difference occurred between the first and second set where average time decreased from 3.926 sec (SE=0.189) to 3.383 sec (SE=0.138). Based on the Tukey pairwise contrast test, the first set was significantly different from sets two through five (p < .0001). The second set was also significantly different from sets four and five (p <.0001). Since the greatest average movement time difference was .543 sec between the first and second set, the first set was removed from the analysis to account for practice (2-3 = .173 seconds, 3-4 = .116 seconds, 4-5 = .081 seconds).

There were significant main effects of AA ($F(7,182)$ = 2.999, $p$ = .005), and TD ($F(7,18170)$ = 163.247, $p$ < .0001), and a significant interaction between AA × TD ($F(49,18170)$ = 4.342, $p$ < .0001) for movement time. PA had no significant effect on movement time ($F(2,182)$ = .879, $p$ = .417). These significant effects are plotted in Figure 7. Movement time increased when the AA changed counterclockwise from 0° to 135°. Similarly, movement time increased when AA changed counterclockwise from 0° to 225°. A distinct drop-in movement time was observed at the 180° AA. A pairwise Tukey test determined that both the 135° ($p$ = .0301) and 225° ($p$ = .020) AAs yielded significantly greater movement times than for the 0° direction. The movement times for an AA of 135° and 225° were 47.2% and 48.7% greater than for the 0° direction, respectively.

Targets located on the diagonals, shown in Figure 7(b), took significantly more time to acquire than targets located in the horizontal or vertical directions, which was indicated by groupings of a pairwise Tukey test. Within these groups, there were no effects of TD on movement time. The interaction between AA × TD is shown in Figure 7(c) in which movement time was segmented by AA.

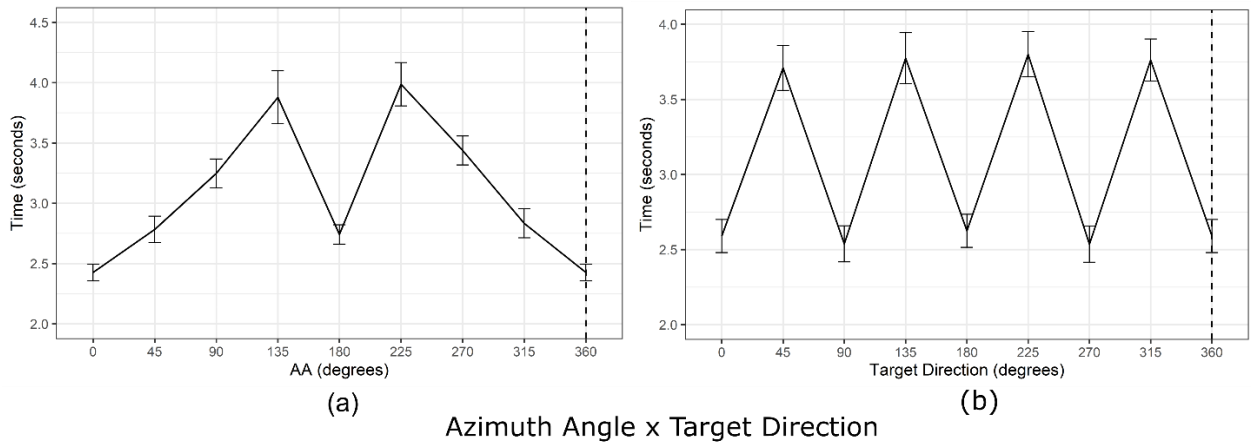

(a)  (b)

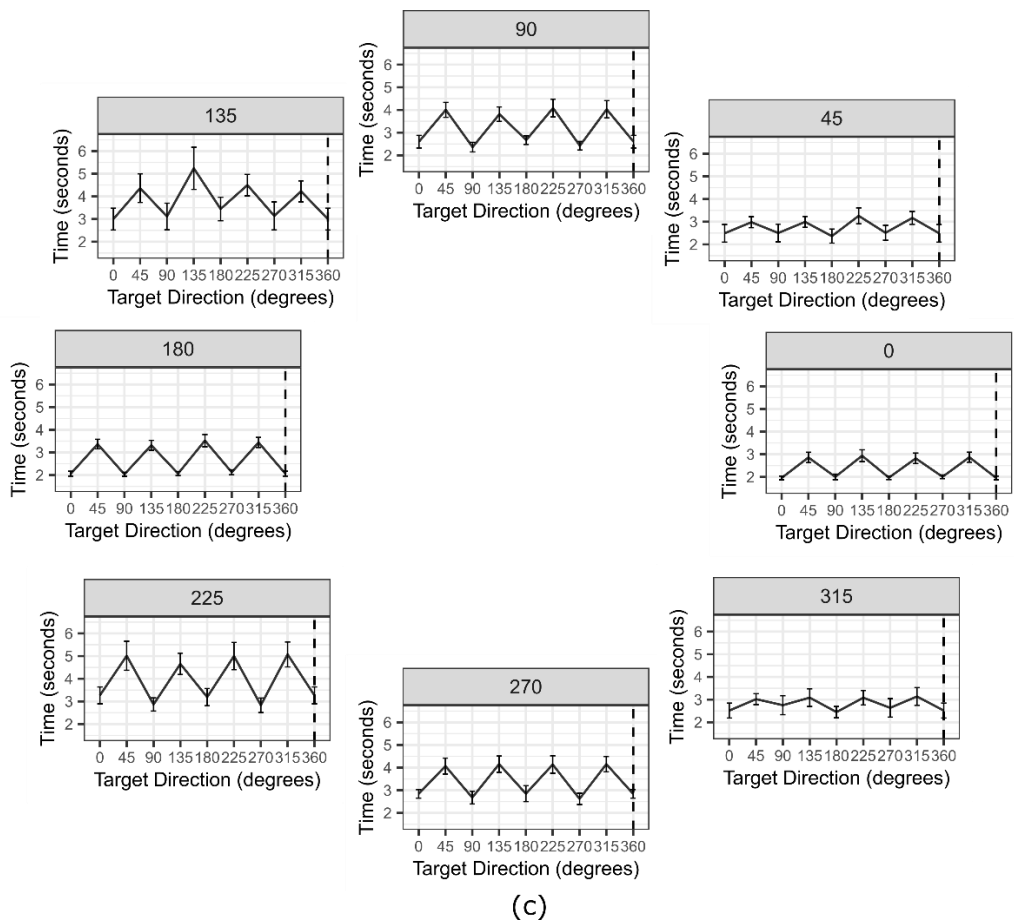

(c)

**Figure 7.** Plots of average time against significant main and interaction effects. (a) Average movement time (±SE) versus AA. (b) Average movement time (±SE) versus the target direction. (c) Interaction between AA and target direction. Note: Target direction of 0° is equivalent to 360° and plotted to display differences for a full rotation of viewpoints.

An additional linear mixed effect model was created to verify Fitts' law held true in the experiment. The independent variables included were azimuthal angle, index of difficulty, and subject. To account for variance, subjects were included as a random intercept in the model. The intercepts and slopes for each azimuthal angle are shown in Figure 8 (Conditional $R^2$=.58).

The AA slope increased counterclockwise from 0° to 135°, which was also seen in the movement time analysis. At 180° there was a drop in the slope and from 225° to 315° the slope decreased. In addition, there were small differences between the intercepts. However, as the index of difficulty increased the more difficult AAs became increasingly difficult. Therefore, Fitts' law showed more difficult tasks became more difficult for more complex viewpoints.

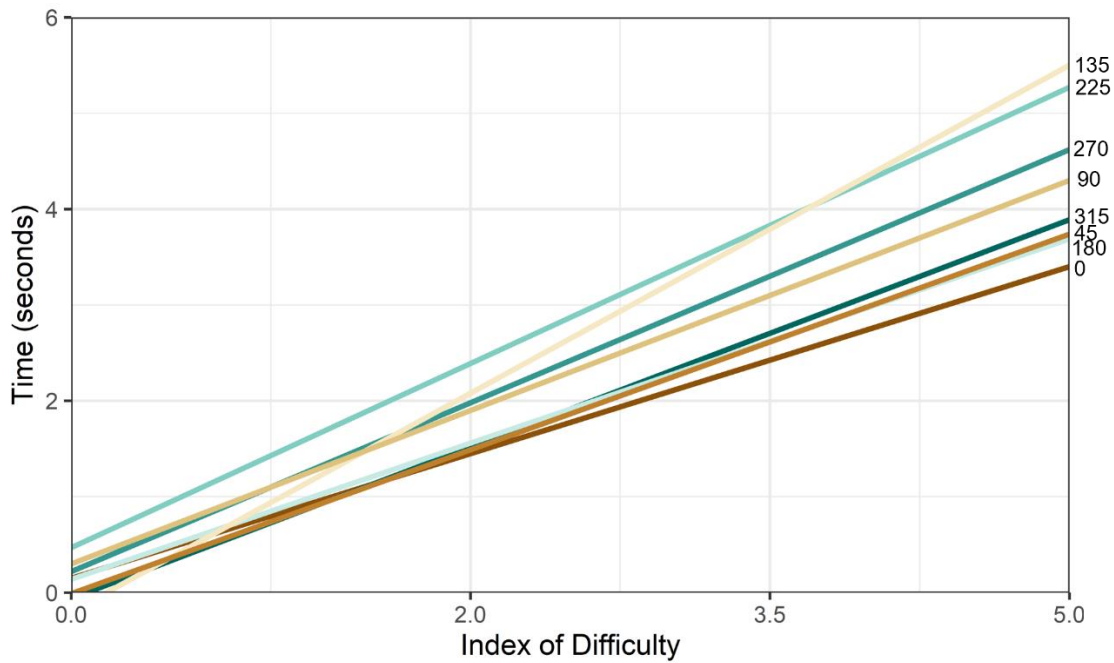

**Figure 8.** Plot of Fitts' law movement time versus index of difficulty for each AA.

*Movement Distance*

Similar learning effects were found for movement distance as for movement time. The set number was statistically significant (F(4,22844) = 28.50, p<.0001). The Tukey test revealed that the first set was significantly different from sets two through five. Since the results aligned with the movement time data, set one was removed from the analysis. Similar results were observed relating the effects of viewpoint on cursor movement distance. The main effects of AA ($F(7,182)$ = 6.852, $p < .0001$) and TD ($F(7,18170) = 24.165$, $p < .0001$) and the interaction between AA × TD ($F(49,18170) = 1.474$, $p < .017$) were also statistically significant. The main effect of PA was not significant ($F(2,182) = 2.554$), $p = .081$) and the interaction between PA x TD was not significant ($F(14,18170) = .627$), $p = .844$). Main effects plots for AA and TD are shown in Figures 9(a) and 9(b), respectively. Analogous relationships can be seen between the main effects for movement time and distance. Movement distance trended upward when AA stepped clockwise from 0° towards 135°, and counterclockwise from 0° towards 225° and then fell dramatically for the 180° angle.

The pairwise Tukey test found several significant differences between AAs 135° and 225° and better performing AAs. The movement time for an AA of 135° was between 14.1-28.4% larger than a better performing AA. Similarly, an AA of 225°, was between 12.1-23.5% larger. The largest differences are seen between the difficult AAs of 135° and 225° and an AA 0° (p<.0001). The distinguishable zig-zag pattern for the target direction effect was observed, whereby the subjects traversed longer distances when pursuing targets in the diagonal locations.

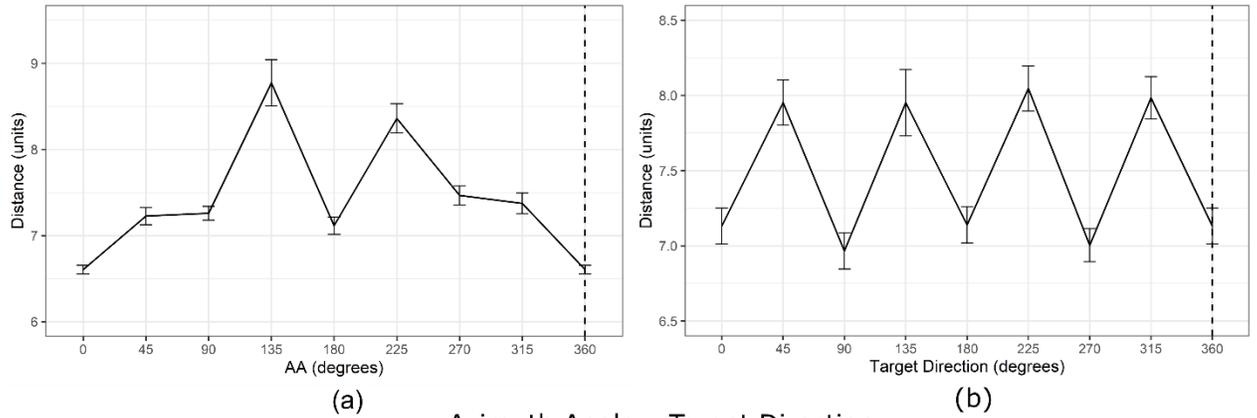
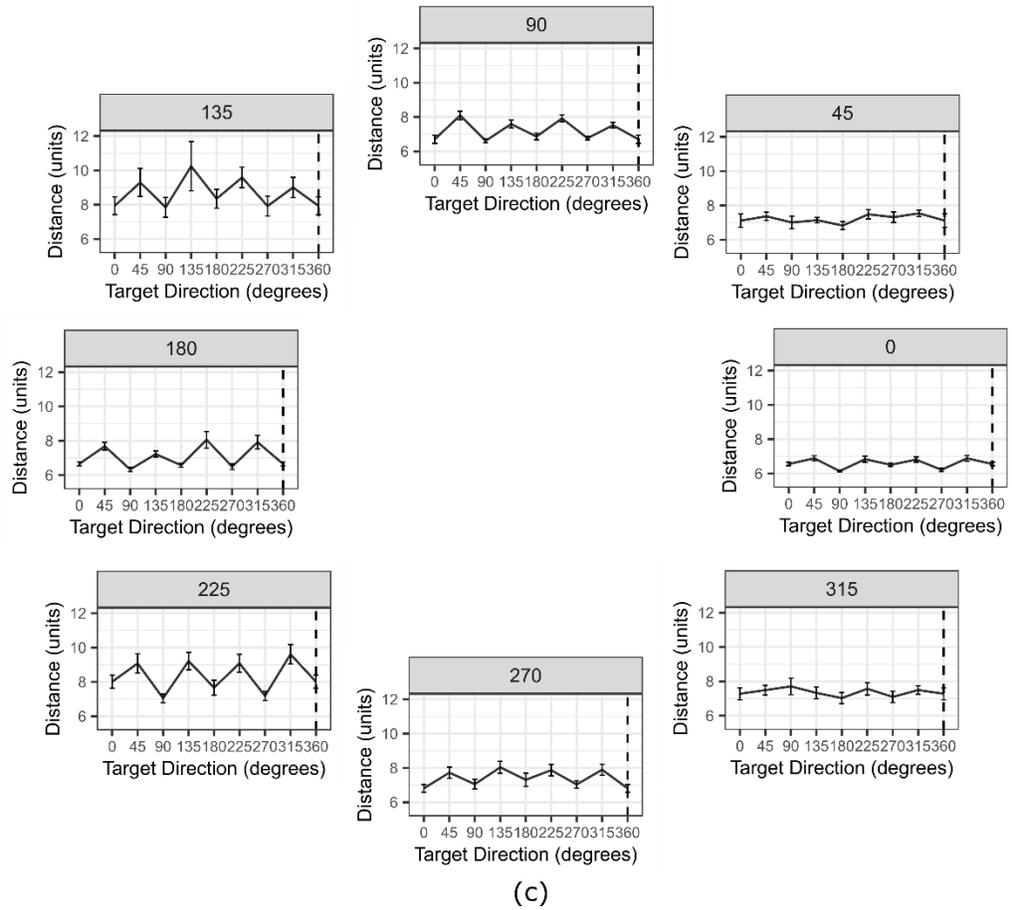

**Figure 9.** Plots of average distance against significant main and interaction effects. (a) Average distance (±SE) versus AA. (b) Average distance (±SE) versus the target direction. (c) Interaction between AA and target direction. Note: Target direction of 0° is equivalent to 360° and plotted to display differences for a full rotation of viewpoints.

## Experiment 2: Target Acquisition for a Changing Viewpoint

*Movement Time*

Similar effects of learning were found in the sets of experiment two ($F(2,21360) = 321.43$, $p < .0001$). A pairwise Tukey test showed each set was significantly different from each other ($p < .0001$). The first set was .69 sec greater than set two and one sec larger than set three (set one = 5.13 sec (SE=.0395), set two = 4.44 sec (SE=.0348), and set three = 4.11 sec (SE=.0266)). Set one was removed to account for learning.

A linear mixed effect model was used to evaluate the effect of previous AA, current AA, and magnitude of transition on movement time. Subjects were considered a random intercept in the model, for subject differences. The magnitude of transition is defined as the number of degrees between the previous AA and current AA. Previous AA was found to have no effect on performance ($F(7,14195) = 1.4192$, p=.234). Current AA did have an effect on performance ($F(7,14195)=160.692$, $p < .0001$). The magnitude of transition was not significant ($F(7,14196)=1.36$, $p =.216$). Orthogonal transitions of 0°, 90°, and -90° had the least influence on performance time. A Welch Two Sample t-test was conducted between Experiment 1 and Experiment 2 to see if dynamic viewpoints had an effect on movement time. A subset of data was used from experiment one, so the conditions in each experiment matched. Even though the magnitude of transition wasn't significant, movement time was significantly greater when there was a change t(1157.5) = 2.02, p = .0438). The second experiment had a mean that was 2% higher or .0594 seconds greater.

*Reaction Time*

Similar learning effects were found for reaction time. Set was significant ($F(2,21360) = 35.55$, $p < .0001$) and each set was significantly different from each other. A linear mixed effect model was computed for the dependent variable of reaction time. The independent variables were previous AA, current AA, and magnitude of transition. Subject was considered a random intercept in the model, to account for the differences between subjects. Previous AA did not have a significant effect on reaction time ($F(7,14195) = 1.743$, $p = .187$). Current AA had a significant effect on reaction time ($F(7, 14195) = 29.178$, $p < .0001$). The magnitude of transition had a significant effect on reaction time ($F(7,14195) = 2.102$, $p = .039$) were found. The effects of transition on reaction time are shown in Figure 10. The Tukey test revealed a transition of -45° was significantly 7.2% greater than a transition of -90° ($p = .047$).

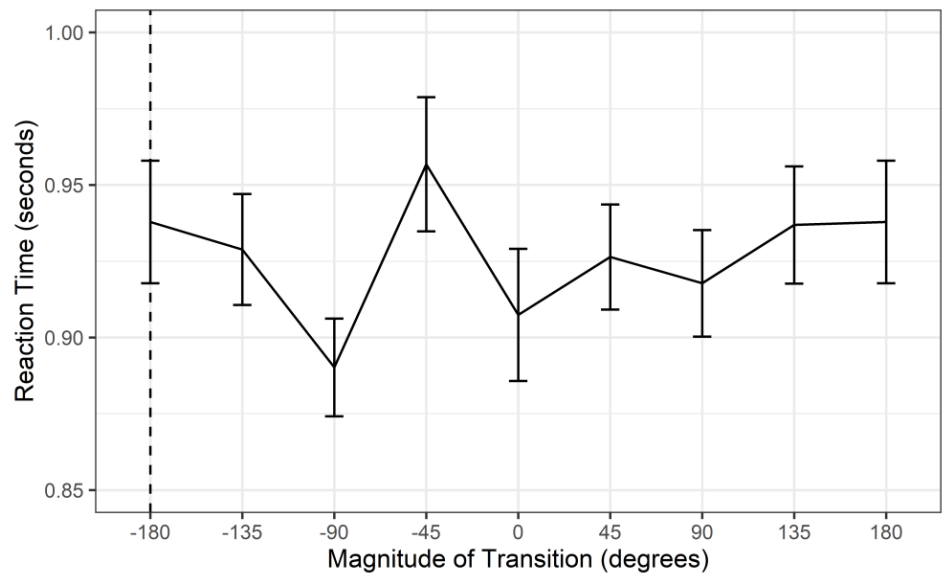

**Figure 10.** Plot of average reaction time against significant main effect of transition.

*Movement Distance*

Learning effects were statistically significant for movement, similar to the dependent variables of time ($F(2,21360) = 10.41$, $p < .0001$). Each set was significantly different from each other. Similar results were found in movement distance shown in Figure 11. Previous AA did not have an effect on movement distance ($F(7,14195) = .03$, $p = .862$), but current AA did have a significant effect ($F(7,14195) = 72.318$, $p < .0001$). Magnitude of transition showed a significant effect on movement distance ($F(7,14196) = 2.133$, $p = .037$). The Tukey test revealed there was a marginally significant difference between a transition of 180° was 3.8% greater than a transition of 0° (*p* = .0673). A Welch Two Sample t-test was used to compare the first and second experiments. The dynamic viewpoints in the second experiment significantly affected movement distance (t(1206.3 = -3.48, p = .0005). The second experiment had a mean that was 4.63% greater.

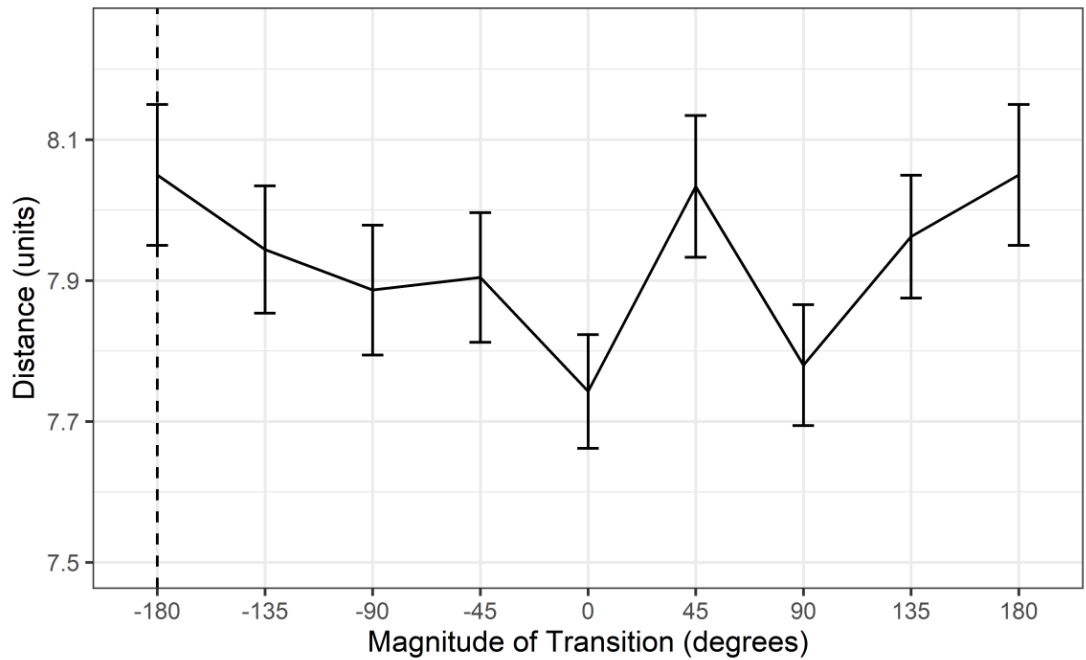

**Figure 11.** Plot of average movement distance against significant main effect of transition.

## Discussion

As hypothesized, viewpoint affected targeting performance. When perceptual-motor misalignments were introduced within the manipulation plane, performance degradation increased as the degree of misalignment increased up to 135° in either direction. However, this trend did not continue up to the maximum possible azimuthal misalignment of 180 °. These results are mostly consistent with the theory proposed by Cunningham (1989) and Bock et al. (2003). They proposed that there is a gradual rotation of an internal mapping to reconcile perceptual-motor misalignments. This rotation of an internal mapping is expected to result in increasingly degraded performance as the distortion increases. However, this gradual adaptation process does not explain the relatively good performance at 180°, which the authors suggest is due to inversion of axes of the manipulation frame. This inversion seemingly produces a lower degradation in performance compared to what would be expected with a gradual adaptation process. The theory also propounds that the gradual adaptation process is only applicable to a certain threshold, theorized to be 90°. Adaptations to misalignments between 90° and 180° may occur due to axes inversion for 180° and a gradual backward rotation to the particular distortion. For example, adaptation to a misalignment of 135° could be attributed to the axes inversion and a 45° backward rotation.

Cunningham (1989) found comparable results from a two-dimensional pointing task study in which azimuthal rotations were mapped from the visual space to the motor space. Poorer targeting performance was observed at 90° and 135° rotations with a noticeable improvement occurring at 180°. More recent work has shown that this pattern holds for visual rotations in laparoscopic training tasks as well (Klein et al., 2015). Our results also indicate a symmetrical pattern of performance for 180° to 360° that can be attributed to a combination of gradual adaptation process and axes inversion. With the exception of 180°, our results seem similar to the findings of Shepard and Metzler (1971) in which the time to mentally rotate an image was linearly related to the angle of rotation. However, perceptual-motor adaptation with continuous visual feedback is thought to depend on a different underlying process compared to a cognitive task such as mental rotation (Mazzoni and Krakauer, 2006). However, recent work such as those by Langsdorf et al. (2021) suggest that mental rotation may have a role in the early phase of perceptual-motor adaptation. Regardless, our results are typical of many previous perceptual-motor adaptation studies, and thus we rely on that literature to explain our findings.

Interestingly, the polar camera angle did not have a significant impact on targeting in this study. Whereas AA is in the same plane as the cursor motion, PA is orthogonal to the plane of cursor motion. Consequently, changes in AA affect viewer orientation with respect to the arrow key directions, while a change in PA is independent of the arrow key orientation. One can conclude that camera location along orthogonal planes of motion will not affect performance. This result has important implications for camera pose selection. The camera can be placed at all locations along the 0° azimuthal arc with minimal performance effects. This greatly increases the space of adequate camera positions by extending the space of easier viewpoints. For applications in which camera location is limited, this result highlights the importance of camera orientation. To illustrate, the two cameras defined in ($\theta$, $\varphi$) coordinates by (225°, 60°) and (45°, -60°) are overlaid but rotated with respect to one another by 180° about the camera viewing axis. When deciding between camera orientation at this location, the results would advise toward selecting the (45°, -60°) pose despite the inverted view because of its smaller azimuthal angle.

Target direction also had a significant effect on both movement time and distance. Targets took more time to acquire in the diagonal directions than in the horizontal and vertical directions. Previous studies of mouse-controlled cursor movement (Whisenand & Emurian, 1996) and head-controlled cursor movement (Radwin et al., 1990) found similar results to the keyboard-controlled tasks in the current study, so these patterns align with continuous controllers. It is also important to consider that diagonal targets require movements in two dimensions to be acquired, whereas horizontal and vertical targets can be acquired in only one dimension. Steps were taken to mitigate the differences between keying diagonal and orthogonal directions, diagonal movements. Although the targets were equally spaced diagonally and orthogonally by their Manhattan distances and that successive keystrokes required the fingers to fully release the key, the need to use two keys may make acquiring diagonal targets inherently require more movement, assuming a constant keying rate. Participants were not instructed on how to utilize their arrow keys such as the number of fingers to use or how to set up the fingers over the keys. Depending on each participant's method, switching between keys may, or may not have had a significant impact on targeting performance. However, the fact that participants needed to switch between keys could have influenced the diagonal movements.

In another mouse-controlled cursor movement study, Thompson et al. (2004) found that vertical targets were more difficult for subjects to acquire than horizontal targets. They hypothesized that this discrepancy could be due to biomechanical differences; horizontal cursor movement with a mouse uses the wrist whereas vertical movement uses the shoulder and upper arm. In this study, however, cursor movement was controlled primarily by the fingers regardless of target direction. This may be why similar results were observed in both the horizontal and vertical directions, as biomechanical effects were not a considerable factor in the current study.

The negative effect of diagonal cursor movement on targeting capability was aggravated by more difficult viewing angles, while controlling for an equivalent number of keystrokes. In other words, the effect of viewing angle was more significant for higher degree-of-freedom movements. In real-world applications, tool and limb movements introduce significantly more complex manipulations with more degrees of freedom. Adding degrees of freedom to a task has been shown to increase movement time for Fitts' law-related tasks (Stoelen & Akin, 2010). Therefore, for tasks which involve more complex movement, the effect of negative viewpoints may be worsened even further, making proper camera pose selection even more important for practical integration. The results of Experiment 1 demonstrated the effect of negative viewpoints. The viewpoints had relatively similar intercepts; however, the transformed spatial mappings required from the negative viewpoints worsened performance at a higher rate than optimal viewpoints. In Experiment 2, a specific viewpoint subset was selected from Experiment 1. Because polar camera angle φ was found to have an insignificant effect on performance in Experiment 1, this angle was held constant at 60° throughout Experiment 2. By collapsing the space of available viewpoints in this insignificant dimension, all possible transitions among the remaining viewpoint subset were able to be presented to each subject in a more reasonable number of trials.

The results for Experiment 2 showed that the previous manipulation perspective had no effect on targeting ability in this task. Performance instead appeared to primarily be a function of the current viewpoint at which one is performing the task. This is contrary to the carryover effects found in several studies (Abeele and Bock, 2001; Neilson and Klein, 2018), where prior exposure to perceptual-motor distortions impacted subsequent exposure. Our results may be due to the short exposure time the participants had to the previous viewpoint, which did not allow them to fully adapt to the displayed distortion. A longer exposure time may allow better adaptation, resulting in carryover effects that improve or degrade performance at a new viewpoint. However, in an actual work environment with dynamic viewpoints, workers are likely to use multiple viewpoints in quick succession. Thus, our results are relevant to the scenarios that we are considering in this paper. However, we acknowledge that these results do not generalize to environments where workers switch between viewpoints after prolonged periods of exposure.

Experiment 1 showed an increase in movement time and distance at AA of 90° and 270°. Although not significantly different, Experiment 2 suggests that orthogonal transitions of -90° or 90° improves performance. A possible explanation for this finding is that subjects in Experiment 2 experienced a changing viewpoint. In a tracking task conducted by Abeele and Bock (2001), results showed worse performance at a rotation of 90°. Further, there was a higher magnitude of errors as rotation angle decreased. In Experiment 2, there was little difference between the magnitude of an increasing and decreasing rotation angle. Performance was defined by root mean square errors in the tracking task (Abeele and Bock, 2001). Differences in experiment results could be from the performance metrics and task. In Abeele and Bock (2001), a joystick was rotated and in the present research the viewpoint was rotated. Despite the differences between Abeele and Bock (2001) and Experiment 2, each movement to a target requires multiple keystrokes.

Band and Miller (2007) confirmed orientation has a significant effect on preparation and reaction time. The results in Experiment 2 show there is less preparation required when the rotation is orthogonal. Smaller preparation time may influence better performance, in terms of movement time and movement distance. Band and Miller (1997) found an increase in reaction time from a transition of 0° to 180° degrees. In Experiment 2, the relationship between orientation and reaction time was not linear. Orthogonal transitions had smaller transition times. In general, results from Experiment 2 increase from 0° to 180° like Band and Miller (2007). However, unlike Band and Miller (2007) the largest reaction times occurred at -45° and 135°.

Therefore, for more complex viewpoints, a transition of -90° or 90° may lead to better task performance. Orthogonal transitions of -90° or 90° influenced better performance than a transition of 180°. This suggests that when a transition of 180° is necessary, it may be better to transition in two steps of 90°. The 180° rule in cinematography supports two steps of 90° because the viewpoint would remain on the same side of the axis, instead of transitioning by a full 180° (Brown, 2011). Ultimately, the performance results from certain types of transitions will influence viewpoint selection when a change in viewpoint is required. The 180° rule can also be used to influence the selections.

The results of this study are applicable to a viewpoint selection algorithm to provide visual feedback while performing manual tasks with an obstructed view. Possible solutions might involve camera positioning that is static, dynamic, or a combination of the two. Regardless of the solution implemented, the results of these experiments would advise towards selecting viewpoints with minimized rotations in the plane of manipulation.

An example of this selection process is illustrated in Figure 12. Starting from the most desirable viewpoint, in this case the viewpoint defined in (θ, φ) coordinates by (0°, 60°), small azimuthal deviations are evaluated as possible viewpoint solutions. The magnitude of allowable deviations is limited by the preset value $\alpha$, which is 1° in this example. If no viable viewpoints are found, the polar angle is decreased slightly, and the search is repeated. This is done until the search terminates, or the polar angle surpasses a preset limit $\varphi_{bound}$. In this example, this value is set to -60° to allow an adequate view of the work surface but may be set as low as -90° depending on the application. If this bound is surpassed, $\alpha$ is increased slightly to now explore viewpoints with a wider azimuthal deviation. This process is repeated iteratively until the best available viewpoint is found, after which the robot is commanded to position the camera accordingly.

Because of the unique circumstances that required the use of this simplified targeting task, there are certain limitations to conclusions that can be drawn from this study. Important factors such as proprioception, muscle fatigue, complex tool manipulation, and high degree-of-freedom movements were not considered. However, this simplified approach allows for a more mechanistic understanding of how transformed spatial mappings can hinder performance when operating through a dissociated visual field and will guide future in-person studies in the realm of visually assisted tasks. The experiments utilized a two-dimensional task, which limits the generalizability to three-dimensional tasks. The results from the two-dimensional task does not necessarily conclude how workers will perform in three-dimensional tasks, but the present research provides potential predictions on performance.

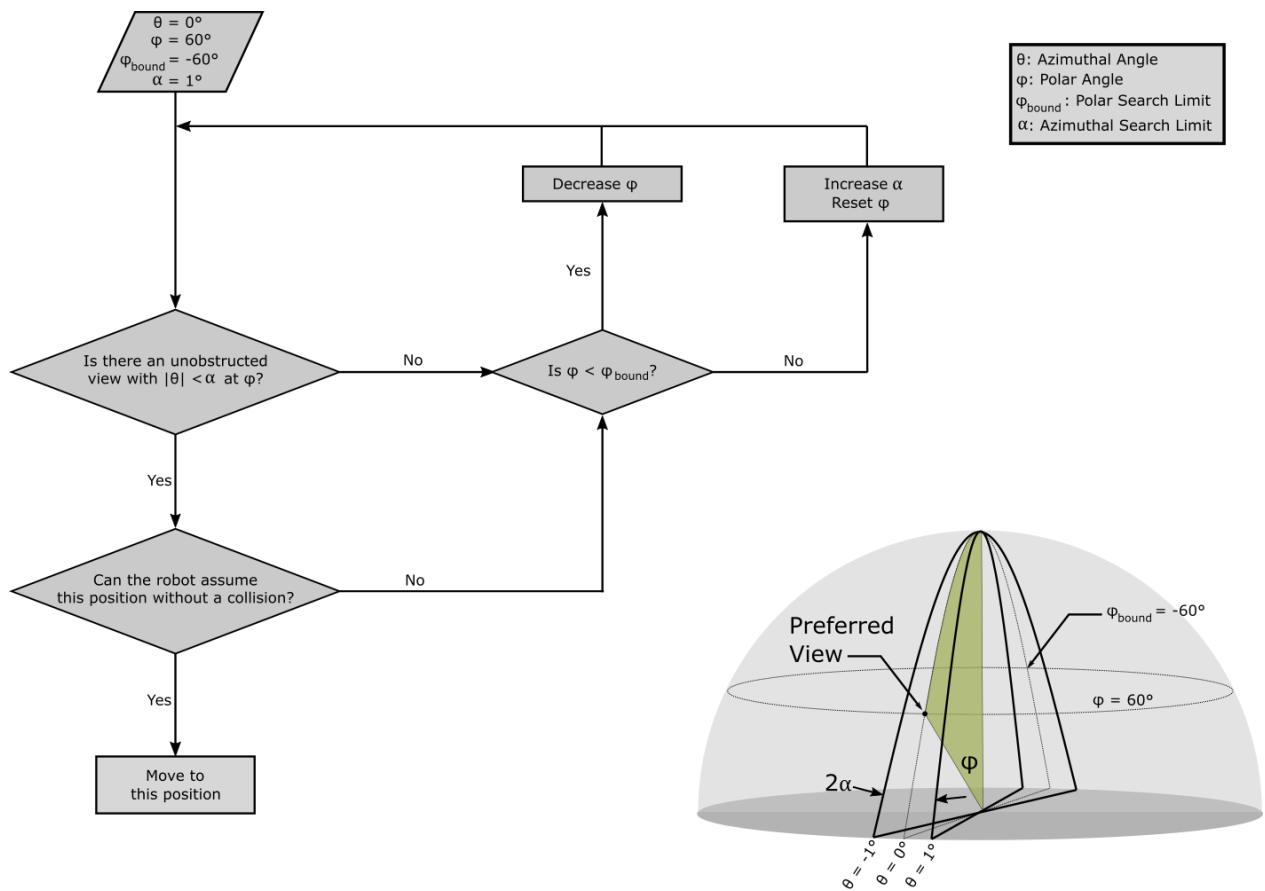

**Figure 12.** Shown above is a simple camera pose selection algorithm informed by the findings of this experiment. Starting from a desirable position, in this case the view with polar coordinates (0°, 60°), viewpoints with azimuth deviations less than $\alpha$ are first looked at as potential solutions. If not, the polar angle is decreased slightly, and this check is repeated. If the bottom polar limit is met, $\alpha$ is increased to broaden the range of acceptable azimuthal angles and the search restarts from the original position. Once a viewpoint has been found that is unobstructed and obtainable without collision, the robot can assume this position with the camera.

In this experiment, the virtual avatar was included to mimic the presence of the worker's extremities when visual feedback is supplied during a manufacturing task. In this way, a visual perception of body location in the virtual space was provided to the subject. Although proprioception was absent in this task, previous research has shown that vision often dominates over proprioception (Chen et al., 2017; Touzalin-Chretien et al., 2010). Therefore, it is unlikely that the effect of proprioception will wash out the main effects found in this study, but it may amplify or dampen the negative effects of alternative viewpoints.

The experiment utilized a two-dimensional task, which limits the generalizability to three-dimensional tasks. The results from the two-dimensional task does not necessarily conclude how workers will perform in three-dimensional tasks. Future research will seek to replicate this experiment in a real-world setting with an analogous targeting task to learn whether these findings are also applicable to physical, three-dimensional tasks. Expert workers may be exposed to different viewpoints over long periods of time, so it is unclear to what extent the results of this study will hold true for them. Future work should address the expert scenario. Further, while the PA did not have an effect in this study, it may become a factor for tasks with higher dimensionality and consequently affect decisioning of the camera placement algorithm. The 180° AA position in particular should be investigated as a potential viewing alternative. Future research will confirm if the benefits seen in this study were an artifact of the simplified control scheme used or they represent a reconcilable inverted output that is also valid in real-world tasks. Additionally, solutions could be created by exploring more efficient work like gradients.

Variables of the visual field presentation modality may also be an important factor when integrating this solution in a manufacturing setting. Hanna et al. (1998) have demonstrated the benefit of minimized display mislocation during endoscopic procedures. For suboptimal viewing angles, however, there may exist cooperative display positions that assist users in performing the associated spatial mapping. In addition to display positioning, future work will investigate various display modalities such as stationary displays, handheld displays, and head-mounted displays to evaluate their efficacy when incorporated.

**Acknowledgements**

This work was supported by a NASA University Leadership Initiative (ULI) grant awarded to UW-Madison and The Boeing Company (Cooperative Agreement # 80NSSC19M0124).

**Key Points**

- Two target acquisition experiments were conducted for a static and changing viewpoint.
- Camera viewpoints up to 135° in both counterclockwise and clockwise directions, linearly diminish psychomotor performance.
- As Fitts' law index of difficulty increased, performance time linearly increased, and more difficult tasks had greater differences between viewpoint azimuth angles.
- When a change in viewpoint occurs, orthogonal changes had the least effect on performance.

DeJong, B. P., Colgate, J. E., & Peshkin, M. A. (2011). Mental Transformations in Human-Robot Interaction. In *Wang X. (eds) Mixed Reality and Human-Robot Interaction. Intelligent Systems, Control and Automation: Science and Engineering* (pp. 35-51). Springer, Dordrecht. doi:10.1007/978-94-007-0582-1_3

Gerges, F. J., Kanazi, G. E., & Jabbour-khoury, S. I. (2006). Anesthesia for laparoscopy: a review. *Journal of Clinical Anesthesia*, *18*(1), 67-78. doi:10.1016/j.jclinane.2005.01.013

Hanna, G. B., Shimi, S. M., & Cuschieri, A. (1998). Task performance in endoscopic surgery is influenced by location of the image display. *Annals of Surgery*, *227*(4), 481-484. doi:10.1097/00000658-199804000-00005

Hiatt, L., & Simmons, R. (2006). Coordinate Frames in Robotic Teleoperation. *2006 IEEE/RSJ International Conference on Intelligent Robots and Systems*, 1712-1719. doi: 10.1109/IROS.2006.282130

Klein, M. I., Riley, M. A., Warm, J. S., & Matthews, G. (2005). Perceived mental workload in an endoscopic surgery simulator. In Proceedings of the Human Factors and Ergonomics Society Annual Meeting (Vol. 49, No. 11, pp. 1014-1018). Sage CA: Los Angeles, CA: SAGE Publications.

Klein, M. I., Wheeler, N. J., & Craig, C. (2015). Sideways Camera Rotations of 90° and 135° Result in Poorer Performance of Laparoscopic Tasks for Novices. *Human Factors*, *57*, 246-261.

Mazzoni, P., & Krakauer, J. W. (2006). An implicit plan overrides an explicit strategy during visuomotor adaptation. *Journal of neuroscience*, *26*(14), 3642-3645.

Menegon, F. A., & Fischer, F. M. (2012). Musculoskeletal reported symptoms among aircraft assembly workers: a multifactorial approach. *Work*, *41*(1), 3738-3745. doi:10.3233/WOR-2012-0088-3738

# Biographies

**Bailey Ramesh** worked on this project as part of his masters research at the University of Wisconsin-Madison.  He has BS and MS degrees in biomedical engineering from the University of Wisconsin-Madison.

**Anna Konstant** is a doctoral student in industrial and systems engineering at the University of Wisconsin-Madison.  She has BS and MS degrees in industrial and systems engineering from Western Michigan University.

**Pragathi Praveena** is a doctoral student in computer science at the University of Wisconsin-Madison.  She has a Btech degree in electrical engineering from the Indian Institute of Technology, Madras, India.

**Michael Gleicher** is a professor of computer science at the University of Wisconsin-Madison where he co-directs both the Visual Computing Laboratory and the Collaborative Robotics Laboratory.  He has a BS degree from Duke University and a Ph. D. in Computer Science from Carnegie Mellon University.

**Bilge Mutlu** is an associate professor of computer science at the University of Wisconsin-Madison where he is the director of the HCI Laboratory. He received a PhD degree from Carnegie Mellon University's Human-Computer Interaction Institute.

**Emmanuel Senft** is a Research Associate in human-robot interaction at the University of Wisconsin-Madison.  He has an MS degree in robotic and autonomous systems from the Swiss Federal Institute of Technology, Lausanne and a PhD from Plymouth University in the UK.

**Michael Zinn** is an associate professor of mechanical engineering at the University of Wisconsin-Madison. He has BS and MS degrees from the Massachusetts Institute of Technology and a PhD from Stanford University in the Stanford Robotics Laboratory.

**Robert G. Radwin** is Duane H. and Dorothy M. Bluemke Professor in the College of Engineering at the University of Wisconsin-Madison, where he advances new methods for research and practice of human factors engineering and occupational ergonomics. He has a BS degree from New York University Polytechnic School of Engineering, and MS and PhD degrees from the University of Michigan.